# Crack-free caustic magnesia-bonded refractory castables

V. C. Miguel[(1)], D. S. Fini[(1)], V. S. Pinto[(1)], M. H. Moreira[(2)], V. C. Pandolfelli[(1,2)], A. P. Luz[(1,2)*]

[(1)] Federal University of São Carlos, Materials Engineering Department,
[(2)] Federal University of São Carlos, Graduate Program in Materials Science and Engineering (PPGCEM)

Rod. Washington Luiz, km 235, São Carlos, SP, 13565-905, Brazil.

*Corresponding author at:
E-mail: anapaula.light@gmail.com or analuz@ufscar.br

**Abstract**

A growing interest in designing high-alumina MgO-bonded refractory castables has been identified in recent years due to the magnesia ability to react: (i) with water at the initial processing stages of these materials (inducing the precipitation of brucite phase) or (ii) with alumina, giving rise to *in situ* $MgAl_2O_4$ generation at high temperatures. Nevertheless, despite the great potential of caustic magnesia to be used as a binder in such systems due to its high reactivity, it is still a challenge to control the hydration reaction rate of this oxide and the negative effects derived from the expansive feature of $Mg(OH)_2$ formation. Thus, this work evaluated the incorporation of different contents of aluminum hydroxyl lactate (AHL) into caustic magnesia-bonded castables, aiming to control the brucite precipitation during the curing and drying steps of the prepared samples, resulting in crack-free refractories. The designed compositions were characterized via flowability, setting behavior, X-ray diffraction, cold flexural strength, porosity, permeability and thermogravimetric measurements. According to the results, instead of $Mg(OH)_2$, hydrotalcite-like phases $[Mg_6Al_2(OH)_{16}(OH)_2.4.5H_2O$ and $Mg_6Al_2(OH)_{16}(CO_3).4H_2O]$ were the main hydrated phases identified in the AHL-containing compositions. The addition of 1.0 wt.% of aluminum hydroxyl lactate to the designed castable proved to be, so far, the best option for this magnesia source, resulting in the development of a crack-free refractory with enhanced properties and greater spalling resistance under heating.

## 1. Introduction

Refractory castables comprise a large group of materials that have evolved, resulting in complex and technical formulations to meet very demanding and harsh industrial applications. The primary features of monolithic compositions are [1–4]: (i) a suitable selection of raw materials (aggregates, fillers, binders, admixtures, and others), leading to a designed particle size distribution, which affects the flowability of the mixture and the resulting density of the consolidated structure; (ii) a proper choice of minor additives (i.e., dispersants, surfactants, setting agents, etc.) that are usually required to control the workability and setting time upon curing; (iii) the binding system, which dictates not only the green mechanical strength and the phase transformations during the first heating treatment, but also the refractory high-temperature properties.

The types of binding agents used in castables have increased in number in the recent decades. Calcium aluminate cements (CACs) are still the main binder for different refractory systems [5–7], but many studies have investigated the advantages and drawbacks associated with the use of hydratable alumina [8,9], colloidal suspensions [10–12], chemical additives [13–15] and even magnesium oxide [16–20], as alternative options for this purpose. The latter has the ability to react with water (in a liquid state or as vapor) giving rise to crystals of the brucite phase [$Mg(OH)_2$], which may fill in the pores available in the consolidated structure of the refractories, inducing the hardening of the molded pieces.

The hydration mechanisms of single crystals and polycrystalline magnesia have been proposed in the literature [17,20–25]. In both cases, it has been reported that the reaction rate strongly depends on the oxide features, environment temperature and admixtures [18–20,26–28]. An important feature of this transformation is that, as the reaction proceeds, the MgO cubic structure is changed to the brucite's hexagonal one, leading to a volumetric expansion and microcracks (the latter, when the expansion is not under control). Moreover, although the hydration rate is slower in liquid water at room temperature, temperatures above 100°C favor vapor formation and, as the steam is in contact with MgO particles, the reaction speeds up favoring the hydroxide crystal growth to take place in a localized, intense and heterogeneous way [21,29]. Consequently, special attention should be given to the magnesia hydration reaction rate during the curing and drying steps of the castables containing such oxide as a binder, as an uncontrolled $Mg(OH)_2$ formation can damage the refractory structure by generating stress gradients associated with brucite expansion in the material [22,30].

The most usual MgO source for a refractory castable matrix is the dead-burnt one, which is commonly obtained by firing magnesite in the temperature range of ~1500-2000°C. Caustic magnesia has the potential to be used as a binder due to its higher reactivity (as it is produced by thermally treating $MgCO_3$ in the range between 600-1300°C), however the faster hydration reaction of this oxide may result in the castable's disintegration during the initial processing steps [20,22,31,32]. Aiming to minimize such a negative effect, previous investigations [18,27,28,33–35] reported that adding organic compounds (i.e., acetate, carboxylic acids, etc.) to MgO-containing compositions might change the morphology and nucleation rate of the formed brucite crystals to better fit their growth in the resulting microstructure. Aluminum lactate is one of these compounds and its action as a chelating agent of $Mg^{2+}$ ions induce the generation of hydrotalcite-like compounds $[Mg_xAl_y(OH)_{2x+2y}](CO_3)_{y/2}.nH_2O$ in magnesia-containing castables [33,36,37], which might help to prevent an excessive brucite formation, especially in compositions comprising caustic magnesia as initial raw material.

To investigate the role of aluminum hydroxyl lactate in inhibiting the drawbacks related to the fast brucite formation, six vibratable caustic MgO-bonded refractory castables were studied in this work. Different contents (0 – 1.25 wt.%) of this organic compound were incorporated into the dry mix of the designed formulations and their preparation and characterization were carried out up to 600°C, which is the temperature range where the hydration drawbacks can damage the product.

## 2. Experimental

*2.1 – Castable compositions*

Six vibratable castable formulations were designed based on Andreasen's model [38] and considering a distribution modulus (*q*) equal to 0.26. Tabular alumina (d < 6 mm, Almatis, Germany), reactive and calcined aluminas (CT3000SG and CL370, Almatis, Germany) and caustic magnesia (QMag 200AR, purity = 98.27%, d < 35 μm, RHI-Magnesita, Brazil) were the main raw materials evaluated in this work, as pointed out in Table 1. Aluminum hydroxyl lactate [(Al(OH)$_3$-x(Lac. Acid)x.nH$_2$O), d < 10 μm, Takiceram M-160P, Taki Chemical Co. Ltd., Japan) was incorporated into the mixtures (0-1.25 wt.%) to adjust the MgO hydration reaction rate and modify the resultant permeability of the designed compositions.

*2.2 – Processing steps and characterization of samples*

The castable's preparation was carried out in a rheometer device [39], where the dry components of the compositions (aluminas, magnesia and dispersant) were initially homogenized for 1 minute, before adding 5.2 wt.% of distilled water to obtain the wet mixtures. After approximately 5 min of wet mixing, the fresh castables were then subjected to vibratable flow tests, according to ASTM C 1445. Curing behavior of the compositions was also monitored via ultrasonic measurements (UltraTest device, IP-8 measuring system, Germany) to follow the propagation velocity of ultrasonic waves in the prepared materials as a function of time and at room temperature (~22°C). Moreover, the prepared mixtures were molded under vibration, cured at 30°C for 24h and dried at 110°C for another 24h.

Table 1

Aiming to understand which phases could be generated in the consolidated structure of the castables, aqueous suspensions containing plain caustic magnesia or a mixture of this oxide with AHL (considering the same MgO:AHL mass ratio as used in the designed refractory composition) were prepared and molded as cylindrical samples (35 mm x 35 mm). After keeping these specimens at 30°C or 110°C for 24h, they were demolded, ground and analyzed via X-ray diffraction tests in D8 Focus equipment (Bruker, Germany), using CuKα radiation [λ = 1.5418 Å], nickel filter, 40 mA, 40 mV, 2θ = 4–80° and a scanning step = 0.02.

Cold mechanical strength measurements were carried out via 3-point bending tests (ASTM C133-97) on cast prismatic castables' samples (150 mm x 25 mm x 25 mm) cured at 30°C for 24h or dried 110°C for 24h and using a universal mechanical device (MTS-810, Material Test System, USA). The apparent porosity of such materials was determined based on the procedure described by ASTM C380-00 and using kerosene as the immersion liquid. A total of five samples was evaluated for each selected testing condition and the presented values are the average result with their respective calculated standard deviation.

Using a device [40], the permeability of the dense castable samples (cylinders with 70 mm of diameter and 25 mm height) was measured after their curing (30°C/24h), drying (110°C/24h) and further thermal treatments at 250°C or 450°C for 5h. The tests were carried out under steady-state conditions and the inlet air pressure (Pi) and volumetric air flow rates (Q) were monitored. Considering these data and Forchheimer's equation (Eq. 1), the permeability constants (Darcian, $k_1$ and non-Darcian, $k_2$) [40] were calculated.

$$\frac{P_i^2 - P_0^2}{2P_0 L} = \frac{\mu}{k_1} V_s + \frac{\rho}{k_2} V_s^2 \tag{1}$$

where $L$ (m) is the sample's thickness, $V_s$ (m/s) is the air velocity, $\mu$ is the air viscosity (1.8 x $10^{-5}$ Pa.s) and $\rho$ is the air density (1.08 kg/m$^3$) at room temperature. Four samples of each selected composition and temperature (110, 250 or 450°C) were analyzed in these tests.

Thermogravimetric analyses were conducted in an electric furnace controlled by a proportional-integral-derivative (PID) system up to a maximum temperature of 600°C and according to different heating rates (5 or 20°C/min). Mass loss during drying was assessed through the normalized parameter $W$ (Eq. 2), which measures the cumulative water content expelled during the heating schedule per total amount of water initially contained in the humid body.

$$W(\%) = 100 x \left( \frac{M_0 - M}{M_0 - M_f} \right) \qquad (2)$$

where $M$ is the instantaneous mass recorded at time $t_i$ during the heating stage, $M_0$ and $M_f$ are the initial and final mass of the tested specimen, respectively. The derivative of the mass loss profiles as a function of time was calculated using the Origin software (version 9, OriginLab, USA).

## 3. Results and discussion

Despite the fact that the same amount of water (5.2 wt.%) was added to the dry mixes during the castables' preparation, some changes in the measured vibratable flow were detected due to the incorporation of aluminum hydroxyl lactate (AHL) into the compositions. According to Fig. 1a, except for CM-0.50AHL, all castables containing the selected additive showed a flow decay when compared to the reference material (flowability changed from 159% to 138%). Previous studies [41] reported that AHL may undergo partial hydrolysis in the evaluated conditions, giving rise to [(R-COO)Al(H$_2$O)$_4$]$^{2+}$ and other species that favor the complexation of magnesium ions [27]. As a result, it was expected that other hydrated phases (in addition to brucite) could be precipitated in the refractories containing this compound. Despite the observed changes, the consistency and rheological behavior of the prepared materials were suitable for a proper casting of the samples.

Figure 1

The setting time of the fresh mixtures was recorded at room temperature (22°C) with the in situ ultrasonic measurements as a function of time and, as few changes could be detected in the obtained velocity profiles, Fig. 1b only highlights the results of the refractories containing 0, 0.5 and 1 wt.% of AHL. As observed, the setting and hardening of the samples took place between 2-4 hours and the CM-1.0AHL formulation presented the highest wave propagation velocity after 24h. Although no cracks were visually detected on the surface of CM-0AHL and CM-0.5AHL samples after the tests, the profile corresponding to these materials indicated that a drop of the measured values occurred after 8 and 14 hours, respectively. Such behavior points out that the generated brucite crystals might have not been properly accommodated in the pores and voids of the resulting microstructure, which induced crack formation in the inner region of the specimens and, consequently, the decrease of the ultrasonic propagation velocity.

According to the X-ray diffraction measurements of MgO-water (MgO = 77 wt.% and $H_2O$ = 23 wt.%) and MgO-AHL-water (MgO = 65.9 wt.%, AHL = 11.0 wt.% and $H_2O$ = 23.1 wt.%) suspensions obtained after their curing (30°C) or drying (110°C) step for 24 hours, periclase and brucite [$Mg(OH)_2$] were the only crystalline phases detected in the additive free mixture (named CM, Fig. 2). Additionally, more intense and well-defined peaks corresponding to the magnesium hydroxide phase was observed when increasing the temperature, which indicates that the hydration reaction proceeded during the drying stage, as water vapor was available in the environment. For the AHL-containing composition, in addition to periclase and brucite, some broad peaks of hydrotalcite-like phases [meixnerite = $Mg_6Al_2(OH)_{16}(OH)_2.4.5H_2O$ and hydrotalcite = $Mg_6Al_2(OH)_{16}(CO_3).4H_2O$] could also be identified due to the interaction of the selected additive with caustic magnesia and water. Both phases present similar diffraction patterns, crystallographic parameters and symmetry, which made it difficult to detect the proportion of each phase. The samples kept at 110°C for 24h presented a small increase in the peaks' intensity of the hydrated phases. Based on these results and some data reported in previous works [33,35–37], the incorporation of aluminum hydroxyl lactate into MgO-containing compositions leads to the generation of hydrotalcite-like phases on the magnesia particle's surface, which inhibits the brucite formation, as pointed out in Fig. 2.

Figure 2

Fig. 3 shows the influence of these phase transformations in the green mechanical strength and apparent porosity of the evaluated refractory castables, after curing (30°C) and drying (110°C) the samples for 24 hours before testing. No cracks were observed on the surface of the reference samples (AHL-free material) after keeping them at 30°C. Nevertheless, as highlighted in Fig. 3c, various cracks were formed in such material during its drying step at 110°C, which indicates that the expansive feature associated with the brucite precipitation and growth of the formed crystals resulted in mechanical stresses in the formed microstructure. This effect was more considerable when the specimens were kept in an environment containing water vapor, as such a condition speeds up the magnesia hydration. Hence, these results highlight how difficult the design of caustic magnesia-bonded refractory compositions is, considering that this oxide has high reactivity with water which favors the generation of a large amount of the brucite phase during the initial processing steps of the castables.

Figure 3

Low flexural strength values were obtained for the castables prepared with 0, 0.25 and 0.5 wt.% of AHL (Fig. 3a). Due to the large number of cracks and the difficulty in handling the dried samples (as some of them were disintegrating and losing small pieces), the apparent porosity for some of these refractories was not evaluated after drying (Fig. 3b). Although CM-0.5AHL specimens still showed some integrity, the presence of cracks and flaws resulted in high porosity levels (110°C) when compared to the compositions containing higher amounts of the selected additive.

Considering the continuous improvement of the green mechanical strength with the increase of the aluminum hydroxyl lactate content added to the caustic magnesia-bonded refractories (Fig. 3a), it was observed that the best performance was achieved when 1.0 wt.% of AHL was used. Most likely, a suitable balance between the brucite crystals and hydrotalcite-like phases generated in the refractory microstructure was obtained in such a condition, which induced a better binding effect among the coarse and fine components of the formulation due to the proper accommodation of the hydrated phases. Temperature also seems to be an important parameter for the enhancement of this property, as a significant increase in the flexural strength was identified for the samples with 0.75 and 1.0 wt.% of AHL after drying at 110°C (Fig. 3a). No cracks were observed on the surface of the dried castables containing 0.75, 1.0 and 1.25 wt.% of AHL. However, the vibratable flow as well as the green mechanical strength of the prepared refractory might be affected when incorporating high amounts of this organic additive (i.e., 1.25 wt.%) into the mixtures. Thus, an optimum content of AHL must be defined for each refractory system, depending on the selected magnesium oxide source.

When comparing both CM-0.75AHL and CM-1.0AHL castables, the latter showed higher flexural strength and apparent porosity values (Fig. 3), which are important aspects when thinking about the development of refractories with enhanced heat up explosion resistance. It is accepted that refractory systems bonded with hydraulic binders (i.e., calcium aluminate cement, hydratable alumina and magnesia) usually present reduced porosity and permeability during their initial processing stages, as a consequence of the precipitation of the hydrated phases that fill in the available pores and voids contained in the consolidated microstructure [5–7]. Although this feature favors the development of materials with improved green mechanical resistance, it might hinder the steam release when they are exposed to fast heating procedures and temperatures above 100°C, which leads to their pressurization and increases their spalling likelihood. Moreover, MgO-containing products are more likely to rehydrate when kept in humid environments (in the presence of steam) for long periods of time, which might also induce damage due to the destructive expansion mechanism associated with the brucite generation [20,23,31]. Therefore, based on these aspects, composition CM-1.0AHL was selected for further characterization tests, as this refractory presented a more promising combination of properties.

The explosive resistance of castables is related not only to the material's porosity but also to the permeability of the microstructure. Thus, additional samples of CM-0AHL and CM-1.0AHL were prepared for the air permeability measurements at room temperature. Knowing that the decomposition of the hydrated compounds (brucite and hydrotalcite-like phases) should change the number of available permeable paths during the drying process of the evaluated refractories, the specimens were dried at 110°C for 24 h or calcined at 250 or 450°C for 5 h, to infer the evolution of this property in this temperature range. Some attempts to measure the permeability of cured samples (kept at 30°C for 24 h and obtained after the demolding step) were also carried out, but due to their well-packed structure, no reliable results were obtained. Additionally, as the reference material (AHL-free formulation) cracked during the drying (see Fig. 3c) and calcining procedures, only the specimens of CM-1.0AHL could be tested.

The calculated permeability constants ($k_1$ and $k_2$) are shown in Table 2. An increase was observed in the $k_1$ and $k_2$ values with the temperature, which can be explained by the water withdrawal and decomposition of the hydrated phases. According to some studies [33,36,37], $Mg_6Al_2(OH)_{16}(OH)_2 \cdot 4.5H_2O$ and $Mg_6Al_2(OH)_{16}(CO_3) \cdot 4H_2O$ compounds present the release of their interlamellar structural water and a simultaneous dehydroxylation and decarbonation of the layered double hydroxide framework around 110-250°C and 300-400°C, respectively. The permeability results obtained after firing at 250 and 450°C were up to two orders of magnitude higher than the ones commonly collected for calcium aluminate cement (CAC)-bonded refractory systems [42,43], when using the same testing conditions. Thus, AHL addition to CM-1.0AHL castable induced the generation of a greater number of permeable paths, which is a key aspect for safe and fast drying.

Table 2

Thermogravimetric analyses of cured (30°C for 24 h) and dried (110°C for 24h) samples of CM-0AHL and CM-1.0AHL samples were carried out to evaluate the spalling resistance of the designed refractories when they were subject to 5°C/min or 20°C/min heating rates. In both heating conditions, the cured CM-0AHL refractory showed a major mass loss close to 135°C (Fig. 4b and 4d), indicating that the generated steam was successfully permeated from the inner region to the external surface of the specimens. However, some cracks were developed for those specimens during the TG measurements, which might have facilitated the water vapor release at such temperatures. Additionally, a second event was identified close to 400°C, which is associated with brucite decomposition. The reference refractory (AHL-free composition) did not have its spalling resistance evaluated after drying because the prepared samples presented a large number of cracks on their surface (see Fig. 3c).

On the other hand, the incorporation of 1.0 wt.% of AHL into the designed castable resulted in a refractory with high explosion resistance and no cracks were generated in such material during the thermal treatments. The main mass loss events took place around 100-250°C with a slower and continuous water release related to the adsorbed and interlamellar water (derived from the hydrotalcite-like phases) withdrawal for the cured samples (Fig. 4a and 4c). After that, the decomposition of $Mg(OH)_2$, $Mg_6Al_2(OH)_{16}(OH)_2 \cdot 4.5H_2O$/ $Mg_6Al_2(OH)_{16}(CO_3) \cdot 4H_2O$ as well as unreacted aluminum hydroxyl lactate [44,45] can explain the broad mass loss peak in the 300-420°C range (Fig. 4b and 4d). When the CM-1.0AHL refractory was subjected to a drying step at 110°C for 24h before the TG measurements, it was observed that the initial mass loss around 80-180°C was suppressed in the obtained profiles (Fig. 4a and 4c). However, the other two peaks at 200-250°C and around 400°C (Fig 4b and 4d) were still present. Moreover, an expected shift of the identified peaks to higher temperatures was detected when increasing the heating rate (changing from 5°C/min to 20°C/min, as shown in Fig. 4b and 4d, respectively).

Figure 4

Therefore, this work presented an interesting new route to produce crack-free caustic magnesia-bonded refractory castables as using a proper amount of aluminum hydroxyl lactate in the designed compositions led to the enhancement of their green mechanical strength, porosity, permeability levels and spalling resistance during their first heating treatment.

## 4. Conclusions

Based on the collected results, aluminum hydroxyl lactate is a suitable compound to help control the caustic magnesia hydration reaction rate during the initial processing stages of refractory castables. The analysis of MgO-containing suspensions indicated that the incorporation of this additive limited the brucite formation and induced the generation of the layered double hydroxides [$Mg_6Al_2(OH)_{16}(OH)_2 \cdot 4.5H_2O$/ $Mg_6Al_2(OH)_{16}(CO_3) \cdot 4H_2O$] on the magnesia particle's surface. Consequently, even when using a very reactive magnesia source (caustic one) as a binder, no visual cracks and damage to the refractory structure were observed during the curing and drying steps, as the reduced amount of formed $Mg(OH)_2$ was well accommodated in the consolidated samples.

Another important aspect is that the layered feature of the hydrotalcite-like phases also plays an important role during the first heating treatment of the caustic MgO-bonded castables, as free-water and interlamellar structural water withdrawal could take place in an extended temperature range (50-300°C), minimizing the explosion risk associated with the steam pressurization in the material. Hence, CM-1.0AHL composition proved to be a good refractory candidate to withstand fast drying conditions (5 and 20°C/min) with greater spalling resistance.

## 5. Acknowledgments

This study was financed in part by the Coordenação de Aperfeiçoamento de Pessoal de Nível Superior – Brasil (CAPES) – Finance Code 001. The authors would like to thank Conselho Nacional de Desenvolvimento Científico e Tecnológico – CNPq (grant number: 303324/2019-8) and Fundação de Amparo a Pesquisa do Estado de São Paulo – FAPESP (grant number: 2019/07996-0) for supporting this work.

[45]  T. Sato, S. Ikoma, F. Ozawa, Thermal decomposition of aluminum hydroxycarboxylates lactate, citrate and tartrate, J. Chem. Tech. Biotechnol. 33 (1983) 415–420.

**Captions for figures**

Fig. 1 – (a) Vibratable flow of the prepared mixtures and (b) evolution of the ultrasound wave velocity propagation in the fresh castables at 22°C.

Fig. 2 – X-ray diffraction profiles of plain caustic magnesia and caustic magnesia + aluminum hydroxyl lactate after their mixing with distilled water at room temperature (22°C), followed by curing (30°C/24h) and drying (110°C/24h) steps. The identified phases in the hydrated samples consisted of B = brucite [$Mg(OH)_2$], M = periclase (MgO) and H = hydrotalcite-like phases [$Mg_6Al_2(OH)_{16}(OH)_2.4.5H_2O$ / $Mg_6Al_2(OH)_{16}(CO_3).4H_2O$].

Fig. 3 – (a) Flexural strength and (b) apparent porosity of the castable samples containing distinct contents of aluminum hydroxyl lactate (AHL). (c) Image of the reference composition (AHL-free refractory) obtained after drying at 110°C/24h, highlighting the crack generation due to the non-accommodation of the brucite phase.

Fig. 4 – Mass loss profiles and their first derivative curves (DTG) obtained for the castables submitted to thermogravimetric measurements with heating rates of (a and b) 5°C/min or (c and d) 20°C/min.

**Figures**

Figure 1

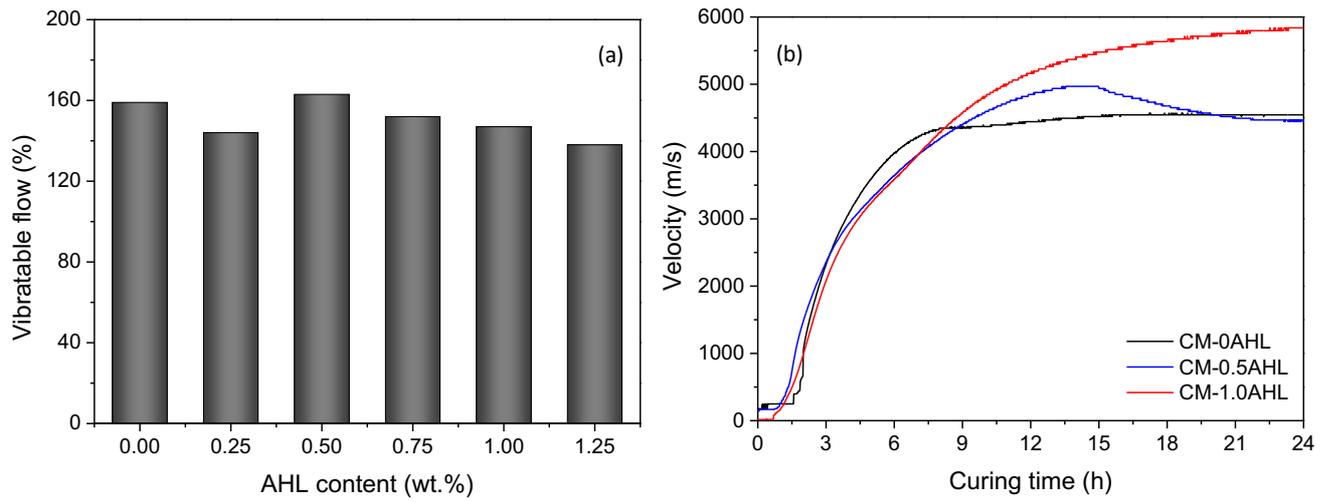

Fig. 1 – (a) Vibratable flow of the prepared mixtures and (b) evolution of the ultrasound wave velocity propagation in the fresh castables at 22°C.

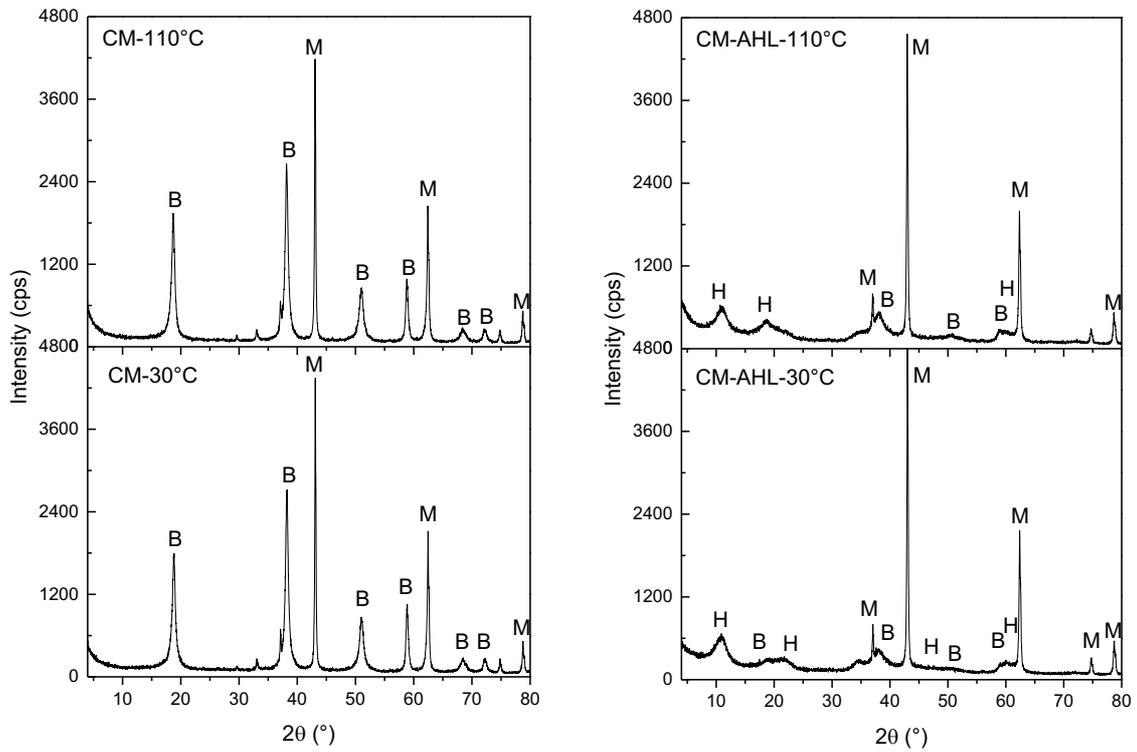

Fig. 2 – X-ray diffraction profiles of plain caustic magnesia and caustic magnesia + aluminum hydroxyl lactate after their mixing with distilled water at room temperature (22°C), followed by curing (30°C/24h) and drying (110°C/24h) steps. The identified phases in the hydrated samples consisted of B = brucite [$Mg(OH)_2$], M = periclase (MgO) and H = hydrotalcite-like phases [$Mg_6Al_2(OH)_{16}(OH)_2 \cdot 4.5H_2O$ / $Mg_6Al_2(OH)_{16}(CO_3) \cdot 4H_2O$].

Figure 3

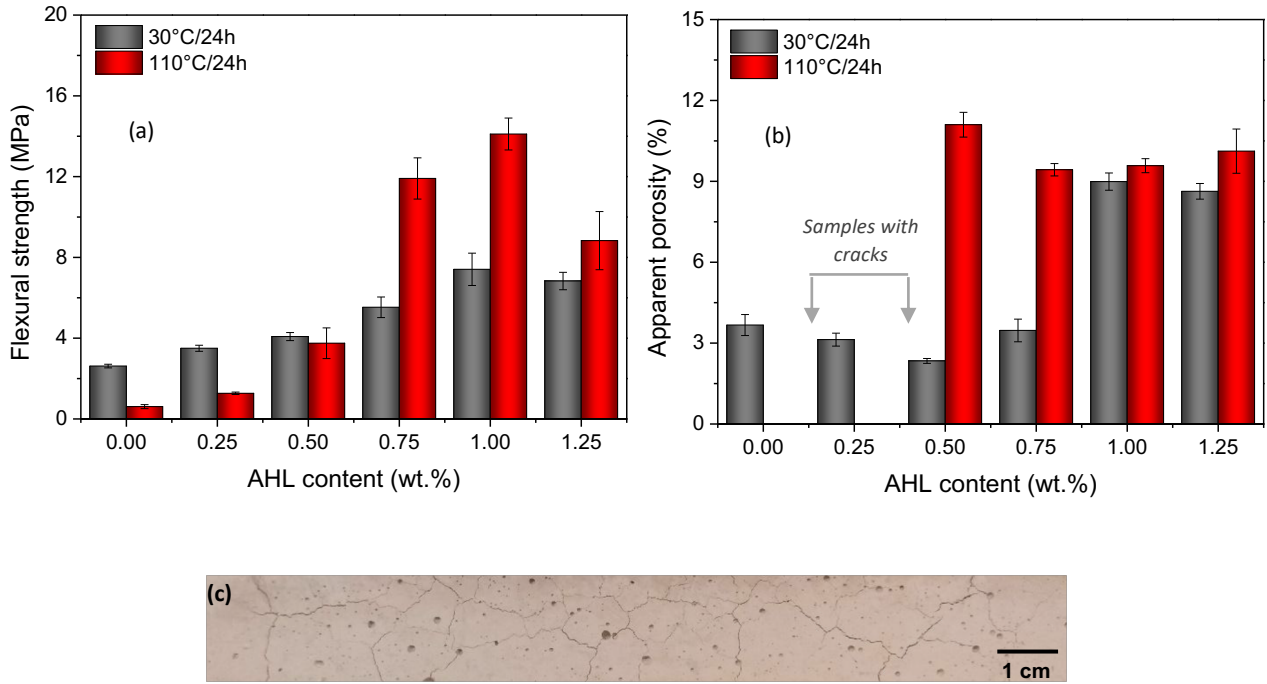

Fig. 3 – (a) Flexural strength and (b) apparent porosity of the castable samples containing distinct contents of aluminum hydroxyl lactate (AHL). (c) Image of the reference composition (AHL-free refractory) obtained after drying at 110°C/24h, highlighting the crack generation due to the non-accommodation of the brucite phase.

Figure 4

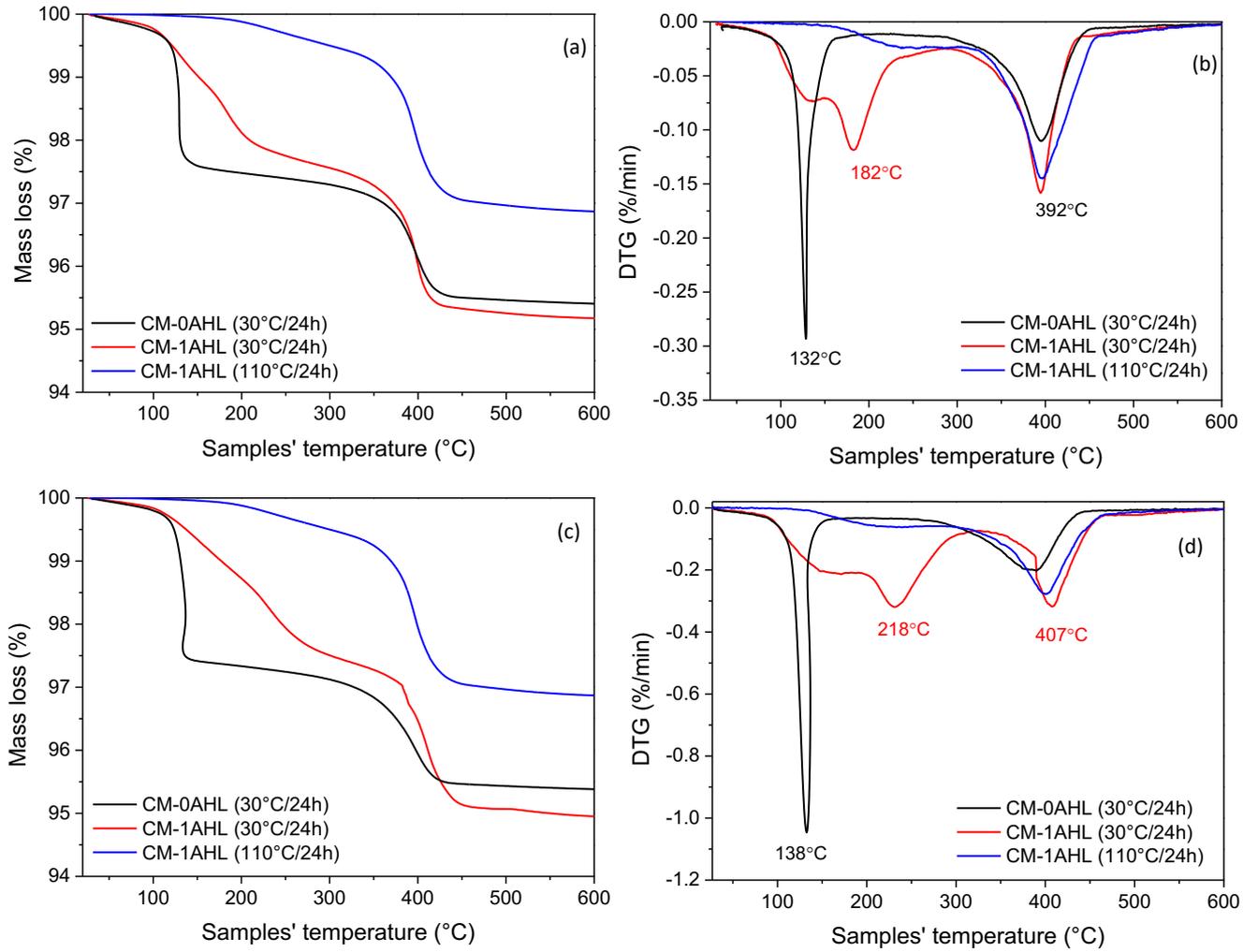

Fig. 4 – Mass loss profiles and their first derivative curves (DTG) obtained for the castables submitted to thermogravimetric measurements with heating rates of (a and b) 5°C/min or (c and d) 20°C/min.

**Tables**

Table 1 – Vibratable castable compositions based on Andreasen's packing model (q = 0.26).

| Raw materials (wt.%) | Compositions | | | | | |
|---|---|---|---|---|---|---|
| | CM-0AHL | CM-0.25AHL | CM-0.5AHL | CM-0.75AHL | CM-1.0AHL | CM-1.25AHL |
| Tabular alumina (d < 6 mm) | 83 | 83 | 83 | 83 | 83 | 83 |
| Reactive alumina (CT3000SG) | 7 | 7 | 7 | 7 | 7 | 7 |
| Calcined alumina (CL370) | 4 | 4 | 4 | 4 | 4 | 4 |
| Caustic MgO (QMag 200AR) | 6 | 6 | 6 | 6 | 6 | 6 |
| Aluminum hydroxyl lactate (AHL) | 0 | 0.25 | 0.5 | 0.75 | 1.0 | 1.25 |
| Dispersant (Castament® FS60) | 0.2 | 0.2 | 0.2 | 0.2 | 0.2 | 0.2 |
| Water content | 5.2 | 5.2 | 5.2 | 5.2 | 5.2 | 5.2 |

Table 2: Permeability constants ($k_1$ and $k_2$) of the CM-1AHL samples after drying at 110°C/24h and calcination at 250 and 450°C for 5h.

| Permeability constants | Drying and calcination conditions | | | | | |
|---|---|---|---|---|---|---|
| | 110°C/24h | | 250°C/5h | | 450°C/5h | |
| | Average | Deviation | Average | Deviation | Average | Deviation |
| $k_1$ (m$^2$) | 9.46 x 10$^{-17}$ | 4.67 x 10$^{-18}$ | 8.07 x 10$^{-17}$ | 7.75 x 10$^{-18}$ | 2.24 x 10$^{-16}$ | 4.39 x 10$^{-18}$ |
| $k_2$ (m) | 5.65 x 10$^{-17}$ | 7.33 x 10$^{-18}$ | 2.99 x 10$^{-16}$ | 4.02 x 10$^{-17}$ | 9.32 x 10$^{-15}$ | 1.25 x 10$^{-16}$ |